\documentclass[]{spie}  %>>> use for US letter paper
\usepackage{amsmath,amsfonts,amssymb}
\usepackage{graphicx}
\usepackage[colorlinks=true, allcolors=blue]{hyperref}
\title{First results of the Test-Bed Telescopes (TBT) project: Cebreros telescope commissioning}

\author[a]{Francisco Oca\~na}
\author[b]{Aitor Ibarra}
\author[b]{Elena Racero}
\author[b]{\'Angel Montero}
\author[c]{Ji\~r\'i Doubek}
\author[b]{Vicente Ruiz}

\affil[a]{Dpto. Astrof\'isica y CC. de la Atm\'osfera, Universidad Complutense de Madrid, Av. Complutense s/n, Madrid, Spain}
\affil[b]{ISDEFE, S.A., Madrid, Spain}
\affil[c]{Iguassu Software Systems, Prague, Czech Republic}

\authorinfo{Further author information: (Send correspondence to F.O.)\\F.O.: E-mail: fog@astrax.fis.ucm.es\\  A.I.: E-mail: aitor.ibarra@sciops.esa.int}

\begin{document} 
\maketitle

\begin{abstract}
The TBT project is being developed under ESA's General Studies and Technology Programme (GSTP), and shall implement a test-bed for the validation of an autonomous optical observing system in a realistic scenario within the Space Situational Awareness (SSA) programme of the European Space Agency (ESA). The goal of the project is to provide two fully robotic telescopes, which will serve as prototypes for development of a future network.

The system consists of two telescopes, one in Spain and the second one in the Southern Hemisphere. The telescope is a fast astrograph with a large Field of View (FoV) of 2.5 x 2.5 square-degrees and a plate scale of 2.2 arcsec/pixel. The tube is mounted on a fast direct-drive mount moving with speed up to 20 degrees per second. The focal plane hosts a 2-port 4K x 4K back-illuminated CCD with readout speeds up to 1MHz per port. All these characteristics ensure good survey performance for transients and fast moving objects. 

Detection software and hardware are optimised for the detection of NEOs and objects in high Earth orbits (objects moving from 0.1-40 arcsec/second). Nominal exposures are in the range from 2 to 30 seconds, depending on the observational strategy. Part of the validation scenario involves the scheduling concept integrated in the robotic operations for both sensors. Every night it takes all the input needed and prepares a schedule following predefined rules allocating tasks for the telescopes. Telescopes are managed by RTS2 control software, that performs the real-time scheduling of the observation and manages all the devices at the observatory \cite{kubanek2010rts2}. At the end of the night the observing systems report astrometric positions and photometry of the objects detected. 

The first telescope was installed in Cebreros Satellite Tracking Station in mid-2015. It is currently in the commissioning phase and we present here the first results of the telescope. We evaluate the site characteristics and the performance of the TBT Cebreros telescope in the different modes and strategies. Average residuals for asteroids are under 0.5 arcsecond, while they are around 1 arcsecond for upper-MEO\footnote{Medium Earth Orbit} and GEO \footnote{geosynchronous satellite } satellites. The survey depth is dimmer than magnitude 18.5 for 30-second exposures with the usual seeing around 4 arcseconds.  
  
\end{abstract}


\section{INTRODUCTION}
\label{sec:intro}  % \label{} allows reference to this section

Space Situational Awareness (SSA) program of European Space Agency foresees the deployment of several robotic telescopes to provide surveillance and tracking services for man-made as well as natural Near-Earth Objects (NEOs). These ground-based optical sensors are very efficient systems to detect and track faint objects  in high altitude orbital regions.

The TBT project is being developed under ESA's General Studies and Technology Programme (GSTP), and shall implement a test-bed for the validation of an autonomous optical observing system in a realistic scenario, consisting of two telescopes located in Spain and a location in the Southern Hemisphere. 

The TBT consortium consists of companies and institutions from the countries funding the project. The prime contractor is the Spanish company ISDEFE, while other members of the consortium are Ixion Aerospace and Industry and the Fabra-ROA Telescope at Montsec \footnote{TFRM http://www.am.ub.edu/bnc/} are members of the consortium. The other partner is IGUASSU \footnote{Iguassu Software Systems (ISS), Prague, Czech Republic} from Czech Republic.
 
The project itself was driven by geographical return and based on COTS components, both for hardware and software. According to the final selection in the Critical Design Review, the operation modes and strategies design were driven by the elements finally purchased.

\begin{figure} [ht]
\begin{center}
\begin{tabular}{c} %% tabular useful for creating an array of images 
\includegraphics[height=5cm]{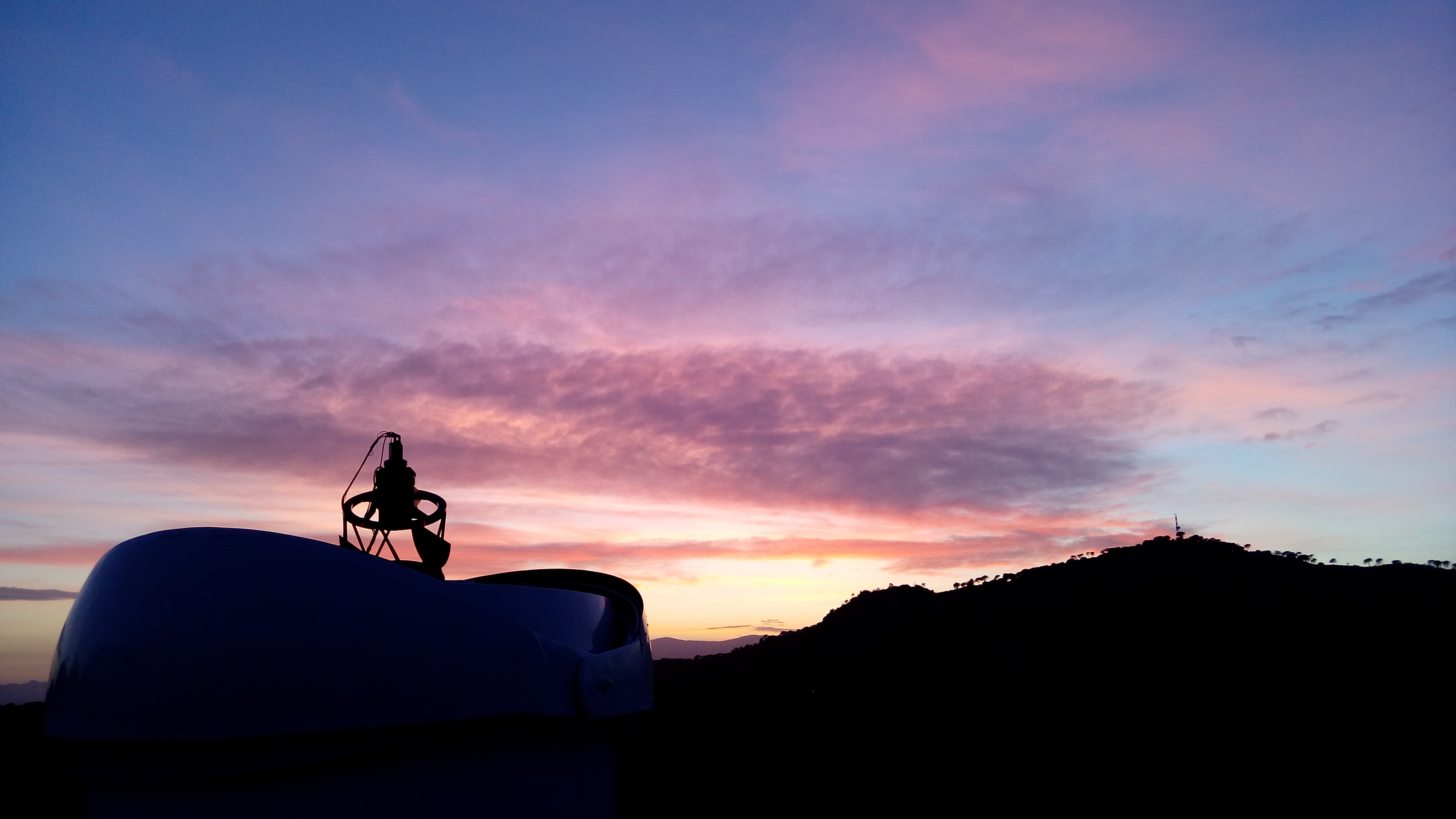}
\end{tabular}
\end{center}
\caption[example] 
 { \label{fig:view} View of the Cebreros TBT Observatory at sunset. All the telescope equipment, control and processing computers are hosted inside of the 4.2~m diameter clamshell dome.}
\end{figure} 

In Winter 2016 the Cebreros telescope has entered in the commissioning phase, led mainly by ISDEFE as the integrator of all the subsystems. After this phase, the second telescope will be installed and both observatories will be tested as a complete autonomous and robotic system. For this Site Acceptance Test, the telescopes will be monitored from ESOC using the Human-Machine Interface HMI, implemented by Ixion, and programmed by the scheduler TRANSITO, developed by ISDEFE. The images are processed and the results sent automatically by TOTAS\cite{koschny2015teide}, software developed by M. Busch and tailored by IGUASSU for the TBT Project. 

\subsection{Telescope design}

The telescope is a fast (f/2.5) prime-focus astrograph with a 3-lens Wynne corrector designed by Dr.~V.~Yu~Terebizh (Crimean Astrophyical Observatory). It has a 56-cm diameter mirror and a resulting unvignetted angular FoV of 3.5 degrees in diameter.

\begin{figure} [ht]
\begin{center}
\begin{tabular}{c} %% tabular useful for creating an array of images 
\includegraphics[height=6cm]{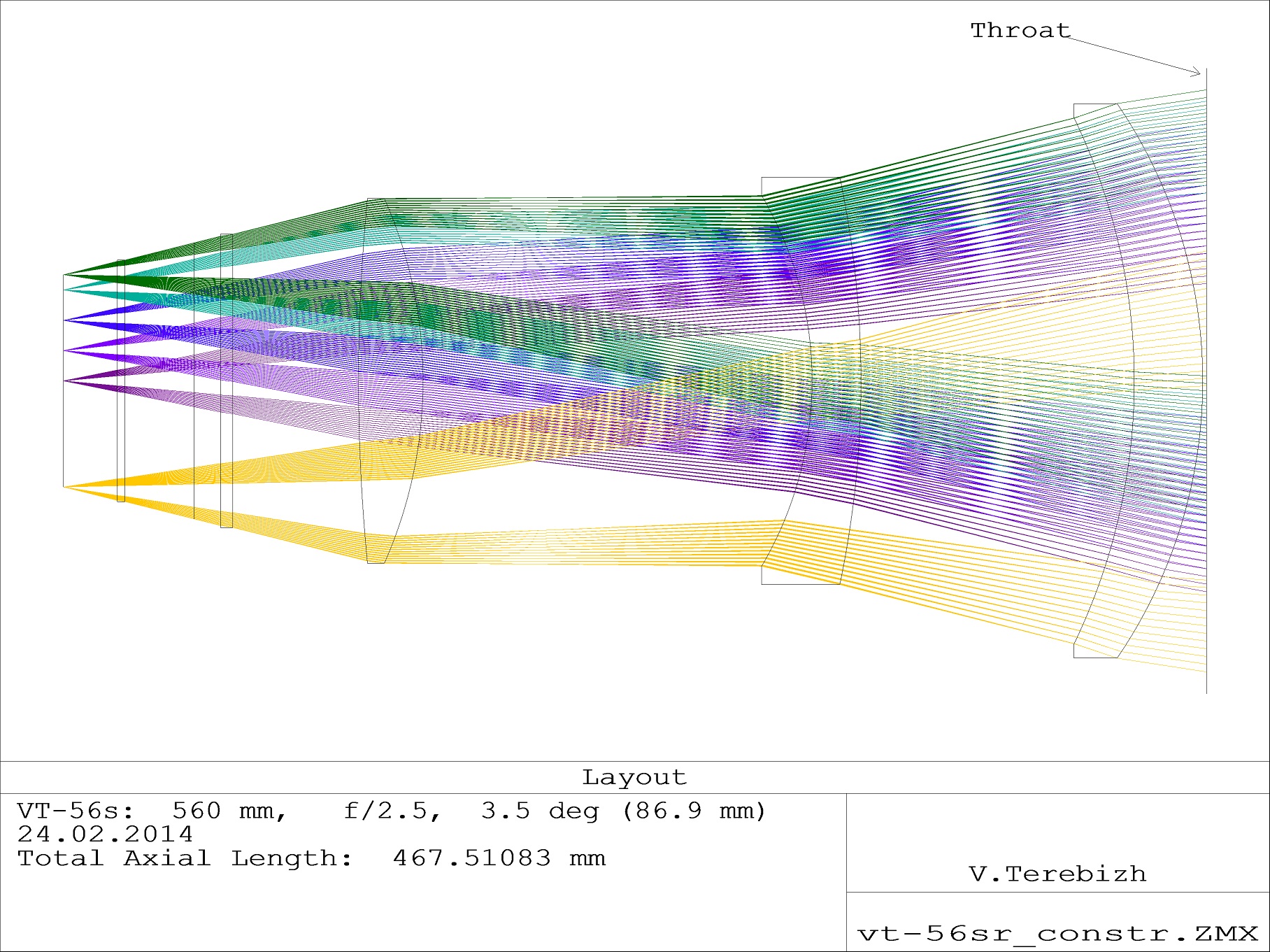}
\end{tabular}
\end{center}
\caption[Optical layout] 
%>>>> use \label inside caption to get Fig. number with \ref{}
 { \label{fig:layout} Detail optical layout of the VT56s design from V.Yu. Terebizh for the TBT telescopes. From the first lens of the Wynne corrector (right) to the focal plane (left). It includes the shutter, filter and the camera window.}
\end{figure}

\begin{figure} [ht]
\begin{center}
\begin{tabular}{c} %% tabular useful for creating an array of images 
\includegraphics[height=6cm]{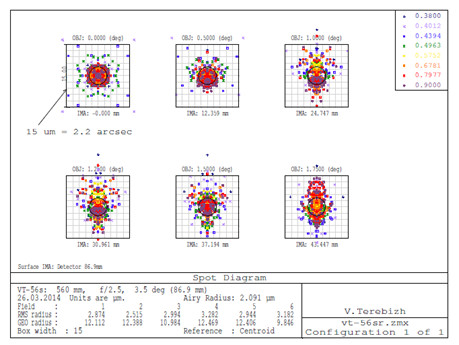}
\end{tabular}
\end{center}
\caption[Spot diagram] 
%>>>> use \label inside caption to get Fig. number with \ref{}
 { \label{fig:spot} Spot diagram in the integral light between 380~nm and 900~nm. Star images correspond to the following field angles (in degrees): 0, 0.5, 1.0, 1.25, 1.5 and 1.75. Box size is 15 microns (2.2\"). Diffraction Airy spot is shown by the circle.}
\end{figure} 

Mirror was polished by LOMO \footnote{Leningrad Optical Mechanical Association, Saint Petersburg, Russia} and Wynne corrector manufactured by TEC \footnote{Telescope Engineering Company, Colorado, USA}. The tube, mount and integration of the four of them were done by APM Telescopes \footnote{Amateur- und Pr\"azisionsoptik Mechanik, Germany}.

\begin{table}[ht]
\caption{Optical parameters and performance of the VT56s design from V.Yu. Terebizh for the TBT telescopes.} 
\label{tab:optical}
\begin{center}       
\begin{tabular}{|l|c|} 
\hline
\rule[-1ex]{0pt}{3.5ex}  Entrance pupil diameter & 560 mm  \\
\hline
\rule[-1ex]{0pt}{3.5ex}  Focal lenght & 1415.5 mm  \\
\hline
\rule[-1ex]{0pt}{3.5ex}  Image space f\# & 2.53 \\
\hline
\rule[-1ex]{0pt}{3.5ex}  Plate scale  & 6.86 $\mu$m/arcsec \\
\hline
\rule[-1ex]{0pt}{3.5ex}  Angular field of view & 3.5 deg \\
\hline 
\rule[-1ex]{0pt}{3.5ex}  Linear obscuration & 0.41 \\
\hline
\rule[-1ex]{0pt}{3.5ex}  Effective aperture diameter & 511 mm \\
\hline
\rule[-1ex]{0pt}{3.5ex}  Back focal lenght & 121 mm \\
\hline
\rule[-1ex]{0pt}{3.5ex}  $D_{80}$ (from center to edge) & 9.1 – 11.4 $\mu$m  (~1.3\" ~- 1.7\"~) \\
\hline
\end{tabular}
\end{center}
\end{table}

\subsection{Mount and pier}

The mount is an APM GE-300 manufactured by Michael Knopf. It is able to load up to 300~kg, although our tube and accessories are significantly lighter. As the telescope is hosted in a clamshell dome, this mount was intentionally selected to perform better while tracking with moderated winds. The mount can move up to 20 degrees per second, but is limited at 12 degrees per second for TBT to increase slewing performance. This speed allows re-positioning in 2 seconds for neighbouring fields and maximum 20 seconds between any two locations in the sky.

TBT telescopes are controlled by Sitech \footnote {Sidereal Technology, Oregon, USA} Telescope Control Systems. The mount has Renishaw absolute encoders with a resolution better than a tenth of arsecond. The servo controller system actuates over two direct driver motors in a close-loop at 2000~Hz. It has proportional-integral-derivative (PID) controller  which has different slewing and tracking modes. These PID parameters were tuned for the requirements of the project.

The mount is set on a pier that is tilted according to the latitude of the site. This design allows the movement of the tube under the mount, removing the collision risk. This position is only used to point to the flatscreen hanging on the wall of the dome.

\subsection{CCD Camera and shutter}

TBT telescopes are equipped with 800S Spectral Instruments cameras. This model has two 16-bit outputs, with readout values between 500~KHz and 1~MHz. Each camera hosts an E2V 230-84 chip with a midband astronomical coating, with a quantum efficiency QE peak \textgreater 95\%. It is a 4K~x~4K chip with 15-micron pixels, covering up to 61mm x 61mm.  The resulting plate scale is 2.18 arcseconds per pixel, and the FoV is 2.5 degrees by 2.5 degrees. 

The camera reads the image from these chips in 12 seconds in the fastest mode. The read-out noise is less than 16 electrons for this mode, while the dark noise is always under 0.2 electrons per second and pixel, at -20 degrees Celsius. For the 500~KHz readout speed, the readout time is 22~s, but the readout noise is under 8 electrons. Linearity and CTE are better than 99\% and 0.999999 respectively. 

For survey the use of binning could reduce the readout time to 2 seconds and thus matching it with the repositioning dead time of the telescope, leading to more than 350 square degrees surveyed per hour (for 30 second exposure,magnitude $\sim$ 19 depth). For follow-up the use of windowing could also reduce the read-out time under 1 second, that is interesting for fast objects. 

% ISDEFE performed a relative spectral response for both of the chips.

The optical beam has an 80-mm shutter manufactured by BS \footnote{Bonn Shutter, Bonn, Germany}, to ensure a 1~ms timing accuracy, with a homogeneity of 0.5\% across the FoV. The shutter needs 150~ms to cover/uncover the whole beam, but the time across the chip can be mapped and corrected thanks to a high level of reproducibility. 

\subsection{Autonomous Emergency System}

In parallel to the nominal systems, the TBT project has the Autonomous Emergency System\cite{ocana2015test,ocana_AES}. It is an Arduino-based system that monitors the environment with redundant sensors (rain, wind, light, power) and closes the dome in case of risk for the operations. It is designed to override the nominal system, though its safety parameters are more relaxed and it only takes control when the nominal system fails to react at the change of conditions. The system relies on two 12~V-batteries that are able to close the dome even at 25\% of charge when there is a shutdown.

\subsection{Ancillary hardware}

Apart from the main systems described here, the TBT project consists of many other elements that we describe briefly in this section. 

The control system takes the weather input from a Davis Vantage Pro 2 weather station. The outdoor sensors are located on a pole next to a nearby building, 4 meters away from the dome. The station is equipped with instruments to measure humidity/temperature (inside and outside the dome), rain and wind speed (gust and averages). 

The calibration of the images includes the use of flat images. The use of dusk/dawn sky images was discarded due to the large gradient present in the FoV. Therefore we use an electroluminiscent flatscreen with a smooth flat spectra covering from 400~nm to 1100~nm. The results are good, but the use of night images is promising. When the telescope archive have enough images, the use of night-sky flats is foreseen. 

In order to avoid hardware failures due to hard shutdowns, all the computers and sensitive electronics rely on a 3000 VA Uninterrupted Power Supply. This UPS is monitored by RTS2 in order to shut down properly all the equipment in case of a prolonged power outage. However the observatory relies on a short-break line, that is never down for more than a few seconds during the automatic start of the station backup diesel generators. 

The camera has a two-stage refrigeration system, where the hot plate of the peltier is cooled down by a 15 degrees Celsius water flux. We use a Thermocube chiller to circulate 2 liters of water per minute. The chiller is controlled and monitored by RTS2. Alarms for low/high temperature and low level of water are active, though in case of temperature increase in the peltier, the camera is automatically shutdown for protection. 

\section{SCHEDULER \& CONTROL SYSTEM}
\label{sec:scheduler}  % \label{} allows reference to this section

According to the high-level architecture design of the TBT project, between the functional layer and the monitoring front-end of the HMI, we have a control layer executing the tasks defined by the scheduler \cite{racero2015towards}. 

\begin{figure} [ht]
\begin{center}
\begin{tabular}{c} %% tabular useful for creating an array of images 
\includegraphics[height=10cm]{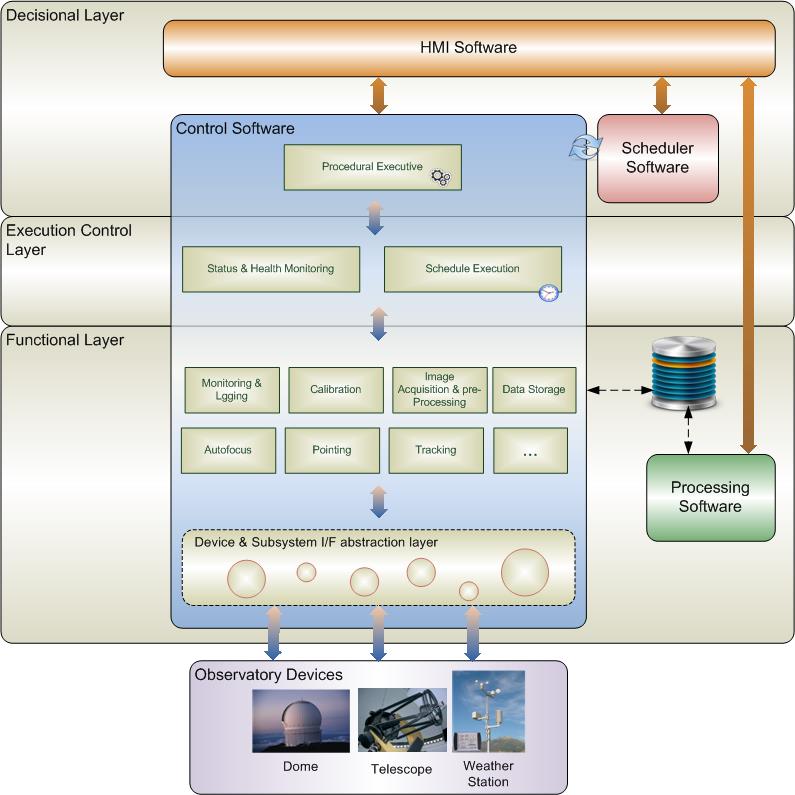}
\end{tabular}
\end{center}
\caption[Architecture] { \label{fig:architecture} Control module top-level architecture.}
\end{figure} 

The scheduling tool (TRANSITO) will implement two general observing strategies for space objects, that is Survey and Follow-Up.  The Survey strategy aims at the detection of new objects for different target population speeds, whereas the Follow-Up goal is the detection of objects in need of revisit.

TRANSITO is  conceived  as  a  high-level  web planning  tool  for  the  manual and automatic planning that will be executed by the telescope control software or RTS2.  This software will  generate  the  Short  Term  Plan  for  the  night,  thus  it  will  be  responsible  for the definition and management of the global strategy for the SSA-TBT observatories, and relies on the RTS2 internal meta-queue system for the real-time response and monitoring of the system. 

RTS2 is is  an  open  source  package  with  the  aim  to  create  a  modular  environment  for complete observatory control\cite{kubanek2004rts} and is oriented to distributed telescope network\cite{kubanek2008rts2} It enables script-driven operation, and permits automatic fast reaction to emerging targets of opportunity.

RTS2 consists of several independent programs. Those programs communicate through RTS2 protocol, acting both as a client and as server, based on TCP/IP connections. Users may intereact with RTS2 either directly, using the RTS2 protocol or through the available built-in bridges: XML-RPC and RTS2-JSON, based on HTTP protocol.

\section{DATA PROCESSING}
\label{sec:data}  

The data processing is done on-site in two different machines. The basic postprocessing is done in the control server, using RTS2 built-in scripts. The images are bias and flat corrected, while the dark is considered negligible. 

The two channels read are merged in one single image, adding all the keywords needed (based mainly in FITS keywords requirement for NEO and SST segments of ESA's SSA project). Astrometric reduction is done using Astrometry.net routine \cite{lang2010astrometry}. 

\subsection{TOTAS}

The TBT Consortium made a trade-off analysis with all the European commercial detection software. In the selection, among other parameters, were taken into account the performance, availability, license and price. This led to the selection of the software from the Teide Observatory Tenerife Asteroid Survey (TOTAS) \cite{koschny2015teide}. This software was written by Matthias Busch for the TOTAS project, a survey performed by ESA's telescope: Optical Ground Station (OGS).

TOTAS was modified for the TBT project in order to detect and measure objects in GEO and MEO orbits. The modification was done by IGUASSU and is based on masking the trailed stars in images taken with on-object tracking. 

\section{TBT CEBREROS TELESCOPE OPERATIONS}
\label{sec:operations}  % \label{} allows reference to this section

According to high-level requirements from ESA, and the hardware and software selected at the CDR, the TBT Project has defined 4 different observation modes that have defined the approach for the Scheduler and Control System.  

The night could be divided up to four different periods, according to the  four different predefined observation strategies performed by the system, that is, Follow-Up of NEOs, Follow-Up of Satellites (GEO, upper-MEO), Survey of NEOs and Survey of Satellites \cite{racero2015toward}. TRANSITO also allows the user to manually schedule a custom observation for the night not predefined by the system. 

\begin{table}[ht]
\caption{Observing nominal modes supported by TRANSITO} 
\label{tab:modes}
\begin{center}       
\begin{tabular}{|l|l|} 
\hline
\rule[-1ex]{0pt}{3.5ex}  Asteroid survey    &  30~s exposure 5x5 pattern 4 revisits, every 15 minutes \\
\hline
\rule[-1ex]{0pt}{3.5ex}  Asteroid follow-up &  Exposure and revisit time depending on object brightness and apparent speed \\
\hline
\rule[-1ex]{0pt}{3.5ex}  Satellites survey  &  2~s exposure, 5 to 8 exposures, depending on the object apparent speed  \\
\hline
\rule[-1ex]{0pt}{3.5ex}  Satellites follow-up & 2~s exposure, 5 to 8 exposures, depending on the object apparent speed \\ 
\hline

\end{tabular}
\end{center}
\end{table}

Part of the validation scenario involves the scheduling concept integrated in the robotic operations for both sensors. At the moment TRANSITO  depends on external services to gather all needed information for the generation of the Short Term Plan for the night. This information will mainly be provided for NEOs by the NEO Confirmation Page and Sky Coverage services from the Minor Planet Centre and by the NEO Priority List from ESA's NEO Coordination Center. For the satellites, ESA's Space Debris Office will feed Tracking Data Messages and Orbital Data Messages with the objects of interest. According to predefined selection parameters, TRANSITO will choose and schedule the objects for the night. RTS2 will receive it and execute them, managing the real-time response to change in environmental conditions. 

\section{COMMISSIONING}
\label{sec:commissioning}  % \label{} allows reference to this section

\subsection{Pointing model and tracking}

TBT project uses the RTS2 built-in pointing model: Gpoint, that stands for GPLed Telescope Pointing model fit \cite{buie2003general,spillar1993wyoming}. This model is intended for equatorial mounts and includes terms to account for non-orthogonality between axis, tube flexure, misalignment of optics and mechanics, zero-point offset, misalignment between instrumental pole and true pole, misalignment to the true meridian, flexure of the declination angle, refraction and hour angle scale error. Some of them are negligible for our system, like the ones associated with periodic gear terms.

The data acquisition is done automatically by a script that makes pointings evenly spaced. The result for TBT Cebreros telescope is a model with 26 arcsecond rms pointing quality, and a maximum value of 79 arcseconds over the whole sky. Those values are 12 and 36 pixels respectively, or in terms of FoV, 1/750\textsuperscript{th} and 1/250\textsuperscript{th}. These values are accurate enough for normal operation of the software, and are a couple of magnitudes better than the requirements. 

We have also evaluated the tracking performance of the telescope. In principle the telescope is going to take short exposures, but we tested it for increasing times. Over 90 minutes (90 exposures of 48s) the range of the values was below 7 arcseconds, with rms 0.9" in right ascension and 1.4" in declination. These values are in good agreement with the model quality, and are within the requirements of the project and expected accuracy with 4-arcsecond seeing. 

For 20-minute tracking (20 exposures of 48 seconds), the range of values is 1.8" in RA and 1.0" in Dec, being the rms 0.41" and 0.26" respectively.  

\subsection{Astrometry and photometry}

Plate solving is performed with Astrometry.net, using a TAN projection third order polynomial, stored in FITS World Coordinate System (WCS) with the Simple Imaging Polynomial (SIP) convention. 

Using stars between magnitude 12 and 18 from UCAC-4 catalogue, we match more than 2000 objects (up to 12000 in crowded galactic plane fields). The average astrometric residuals are under 0.1 arcseconds in both right ascension and declination axis, with no spatial dependence, having a random error distribution across the image. Photometry fitting shows random distribution, with average residual of 0.1 magnitudes in V band.     

\subsection{Asteroid observation}

During the months of commissioning, the Cebreros TBT telescope has performed some survey campaigns. We have analysed the data for 4 fields observed in 4 different nights ( 20 minutes effective time). The average seeing for those nights was 4-4.5 arseconds. We took 324 positions of more than 80 different objects ranging between magnitudes 12 and 19. 

The average total residual was $0.46 \pm 0.42$ arseconds, with less than 7\% of the measurements worst than 1 arcsecond. All of the objects were slow compared to NEO category. The fastest of these 80 objects was around 1 arcsecond per minute.

\begin{figure} [ht]
\begin{center}
\begin{tabular}{c} %% tabular useful for creating an array of images 
\includegraphics[height=5cm]{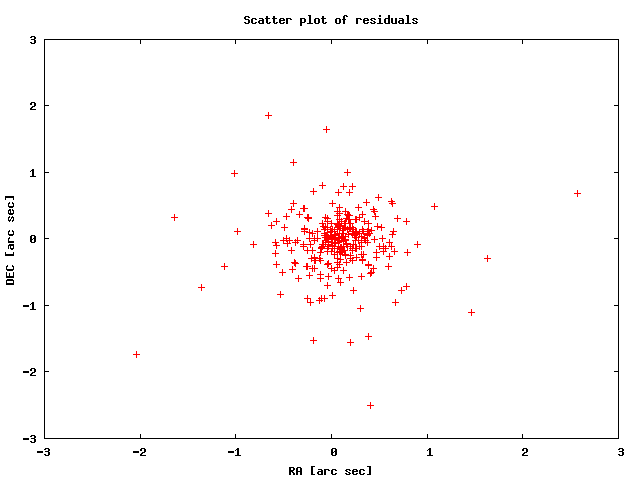} 
\includegraphics[height=5cm]{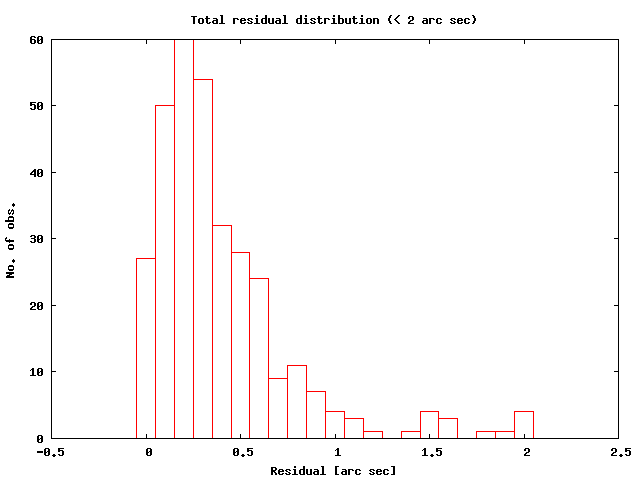}
\end{tabular}
\end{center}
\caption[residual]  { \label{fig:residual} Left: Scatter plot of the residuals of 324 positions of asteroids. All the residuals large than 1 arcsecond are for objects of magnitud 18 and dimmer. Right: Distribution plot of the asteroid observations residuals, with the peak in 0.2 arcseconds. Plots made by the Calculations of residuals service (Fitsblink.net), by Jure Skvar\~c.}
\end{figure} 

The depth of the survey is usually magnitude 19 for exposures of 30 to 60 seconds depending on sky brightness and seeing conditions. 

\subsection{Fast objects residuals}

Fast NEOs and satellite observations are very sensitive to timing uncertainties, consequently part of the efforts of the commissioning has been employed in this topic. The control server is the one stamping the time in the image file. This server is connected to an NTP time server inside ESA's network. We have tested the timing of the images up to the level of uncertainty of the shutter. The shutter has a travel time of 0.15~s across the FoV. 

For NEOs we have observed up to a maximum of 35 arcseconds per minute, with residuals under 1 arcsecond. 

\subsection{Satellites and space debris survey and follow-up}

During the commissioning period, the Cebreros TBT telescope has been performing satellite observations regularly. Most of the observations have been done for satellites in GEO orbital regime reaching magnitude 18 in 5-second exposures. For calibration purposes we observe regularly Galileo, GLONASS and GPS satellites. 

For GEOs (moving around 15 arcseconds per second) the residuals are also better than 1 arcsecond, using the star trails as astrometric reference (measured by our collaborators from the TFRM telescope\cite{montojo2011fabra,fors2013telescope}, that are part of the TBT consortium). The fastest object observed until the moment has been Exomars, moving between 40 and 80 arcseconds per second. The residuals in this case were around 1 arcsecond. 

\subsubsection{Exomars detection}

On March 14\textsuperscript{th}, the European Space Agency launched Exomars. The probe was sent to Mars using different rocket stages. Cebreros TBT telescope was well located to observe Exomars after the last boost. The spacecraft was first imaged at 20:10 UTC, located less than 5000 km above Earth's surface, already with escape velocity. The last image was at 20:16 UTC with the object below 15 degrees of elevation and still moving over 40 arcseconds per second.

\section{Conclusions}
\label{sec:conclusions}  % \label{} allows reference to this section

TBT project will deploy two robotic telescopes to perform SST and NEO observations for the European Space Agency. The first telescope was installed in Cebreros ESTRACK Station and has been successfully commissioned. The second telescope will be installed and commissioned during 2017, after the confirmation of the location at the Southern Hemisphere. 

\acknowledgments % equivalent to \section*{ACKNOWLEDGMENTS}       

The authors would like to thank all the members of the consortium. In special the team of the Telescope Fabra ROA Montsec (TFRM) for valuable discussions for the commissioning of the Cebreros telescope; the support of P. Kub\'anek and S. V\'itek during the installation and commissioning observations.

We are very grateful to all the staff at Cebreros ESTRACK station for their support during the days and nights of the installation and commissioning, especially to Rogelio Marante. 

We also appreciate a lot the feedback received from some ESA staff and contractors involved in the project: Igor Zayer, Gian Maria Pinna, Marco Micheli, Detlef Koschny, Beatriz Jilete and Tim Flohrer.

In memoriam Daniel Ponz.  For us you will be always in the stars.

% References
\bibliography{report} % bibliography data in report.bib
\bibliographystyle{spiebib} % makes bibtex use spiebib.bst

\end{document}